\documentclass[journal]{IEEEtran}
\usepackage{cite}
\ifCLASSINFOpdf
  \usepackage[pdftex]{graphicx}
\else
  \usepackage[dvips]{graphicx}
\fi
\usepackage{amsmath}
\usepackage{amssymb}
\usepackage{url}
\usepackage{import}
\begin{document}
\title{
Quantifying Experimental Edge Plasma Evolution via Multidimensional Adaptive Gaussian Process Regression
}
\author{Abhilash Mathews and Jerry W. Hughes%
        \thanks{This work is supported by the U.S. Department of Energy (DOE) Office of Science under the Fusion Energy Sciences program by contract DE-SC0014264 and the Natural Sciences and Engineering Research Council of Canada (NSERC) by the doctoral postgraduate scholarship (PGS D)}
}
\maketitle
\begin{abstract}
The edge density and temperature of tokamak plasmas are strongly correlated with energy and particle confinement and their quantification is fundamental to understanding edge dynamics. These quantities exhibit behaviours ranging from sharp plasma gradients and fast transient phenomena (e.g. transitions between low and high confinement regimes) to nominal stationary phases. Analysis of experimental edge measurements therefore require robust fitting techniques to capture potentially stiff spatiotemporal evolution. Additionally, fusion plasma diagnostics inevitably involve measurement errors and data analysis requires a statistical framework to accurately quantify uncertainties. This paper outlines a generalized multidimensional adaptive Gaussian process routine capable of automatically handling noisy data and spatiotemporal correlations. We focus on the edge-pedestal region in order to underline advancements in quantifying time-dependent plasma profiles including transport barrier formation on the Alcator C-Mod tokamak.
\end{abstract}
\IEEEpeerreviewmaketitle
\section{INTRODUCTION}
\begin{figure}[h]
\includegraphics[width=0.9\linewidth]{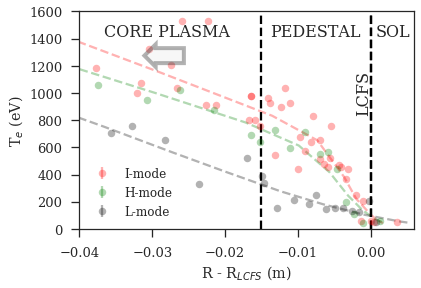} \\
\includegraphics[width=0.9\linewidth]{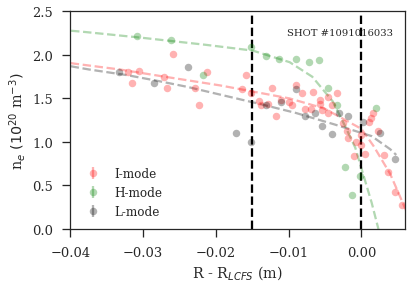}
\caption{Electron density and temperature measurements fit by a modified hyperbolic tangent via nonlinear least squares exhibiting strong---albeit limited in fitted structure---profile variation across confinement regimes in a single discharge. Data and conventional fitting of profiles adapted from \cite{Whyte_2010}.
}
\label{Whyte2010_LHI}
\end{figure}

\IEEEPARstart{T}{he} edge-pedestal of fusion plasmas traverses a vast range of physical scales. At the top of the pedestal are temperatures exceeding the sun's core at nearly $10^7$ K while separated sometimes just centimeters from solid plasma-facing components having operational surface temperatures of $\approx 10^3$ K. This physical span in tokamaks includes some of the strongest temperature gradients observed in the universe and is critical towards overall reactor performance, stability, and fuelling \cite{Groebner_2013,Snyder_2009,Hughes_2018,osti_1567992}. Fundamental plasma properties (e.g. collisionality, ionization rate) consequently vary strongly both spatially and temporally in this region and require capable numerical analysis tools since quantifying the pedestal is essential towards understanding dynamics near the last closed flux surface (LCFS) and across the scrape-off layer (SOL). The advent of scientific machine learning could potentially advance understanding in this region as well, but requires constructing good statistical estimates of partially observed variables from diagnostic measurements. Automatically generating accurate pedestal plasma density and temperature profiles requires sufficiently sophisticated tools to handle large quantities of noisy observations. Current fitting routines in the fusion community involve a range of methods including nonlinear least squares via modified hyperbolic tangent functions \cite{Groebner_2001} such as in Figure \ref{Whyte2010_LHI}, cubic splines \cite{Diallo_2011}, and various Bayesian techniques \cite{Fischer_2010}. Past Gaussian process (GP) regression codes typically fixed kernels with length scale functions already assumed \cite{Chilenski_2015} and generally permitted only one-dimensional scenarios to build radial profiles. This requires filtering or averaging temporal variation in data which can be limiting in the edge especially when analyzing transient phenomena such as spontaneous transitions of confinement regimes and transport barrier formation \cite{Physics__2007, Mathews_2019}. Robustness to capture both mean spatial and temporal variations of edge plasma profiles and associated gradients is therefore sought for improving analysis.
\\
Towards this task a deep multidimensional heteroscedastic GP routine is outlined to provide automated fitting and uncertainty estimates from the Thomson scattering diagnostic on the Alcator C-Mod tokamak. Evolution of Both plasma density and temperature is tracked across the edge-pedestal region with varying length scales extant in experimental data including the formation of both particle and energy transport barriers. This technique has the capability to routinely process thousands of discharges automatically to yield profile statistics and be run across novel experiments. The applied methodology is described in section 2 (with accompanying mathematical proofs in the Appendix), demonstrated on relevant scenarios exhibiting transport barriers from real experiments in section 3, and final summarizing remarks are presented in section 4.

\section{METHODOLOGY}
The technique applied for reconstructing edge-pedestal plasma profiles is an adaptive heteroscedastic multidimensional GP routine. Each of these terms are now individually defined and outlined to introduce the scope of this method. 
\subsection{Gaussian process}
A GP is a supervised learning method capable of solving classification and regression problems. The latter fitting capability via nonlinear probabilistic regression is the main focus of this paper. In particular, the underlying assumption is that the variable (e.g. plasma density) being predicted at a certain location is normally distributed and spatiotemporally correlated with neighbouring points, indicating that partial observations provide information at nearby locations for conditioning future predictions. The function space definition of a GP is that any finite collection of the random variables modelled follow a joint Gaussian distribution, i.e. $(y_i,y_j) \sim \mathcal{N}(\mu,\Sigma)$ and any subset is given by $y_i \sim f({\bf x}_i) + \sigma_n\mathcal{N}(0,I)$ \cite{Rasmussen_2005}, where $\sigma^2_n$ is the noise variance. A GP is specified entirely up to its second-order statistics as denoted by the mean, $\mu({\bf x}_i)$, and covariance, $\Sigma({\bf x}_i,{\bf x}_j)$. Consequently, it can be proven (c.f. Appendix) that conditional predictions at ${\bf x}_b$ based upon observations at ${\bf x}_a$ are analytically given by \cite{Rasmussen_2005}:
\begin{align}
\label{GPR_machinery1}
{\boldsymbol\mu}_{{ \bf y}_b \vert { \bf y}_a} = \boldsymbol\mu_b+\Sigma_{b,a}{\Sigma_{a,a}}^{-1}({\boldsymbol y_a}-\boldsymbol\mu_a)
\end{align}
\begin{align}
\label{GPR_machinery2}
{\Sigma}_{{ \bf y}_b \vert { \bf y}_a} = \Sigma_{b,b}-\Sigma_{b,a}{\Sigma_{a,a}}^{-1}\Sigma_{a,b}
\end{align}
Suitably selecting the kernel function, $k({\bf x}_i,{\bf x}_j)$, composing the full covariance is critical to the GP since it specifies the correlation between any pairs of random variables. Constraining kernels will consequently limit the range of behaviour that can be captured by the GP which may only be physically warranted in certain scenarios. To remain robust to tracking a wide range of spatiotemporal behaviour, an adaptive heteroscedastic kernel is optimized against experimental observations. Applicability depends upon the case, but utilizing alternative distributions (e.g. log-normal) and latent variable transformations can also permit scenarios with non-Gaussian residuals \cite{wang2012gaussian}, although Gaussianity is assumed here.

\subsection{Adaptivity}
GP regression is a nonparametric method without an explicit functional form. Nonparametric in this context means that there are no fixed number of constraining model parameters but instead the fitting routine becomes increasingly constrained as training data increases. Correlations between observed data points are based upon the prescribed covariance function. Various kernels have been proposed to embed this structure ranging from a Gaussian function (for expectedly smooth behaviour) to periodic functions (for expectedly cyclic behaviour) to combinations of multiple kernels (e.g. for automatic relevance determination) \cite{DD_thesis}. Despite the generic regression technique lacking a strictly fixed functional form, the optimized hyperparameters defining the kernel are typically constrained themselves. For example, a standard stationary isotropic Mat\'ern kernel is defined by
\begin{equation}
    k({\bf x}_i,{\bf x}_j) = \sigma^2_k\frac{2^{1-\nu}}{\Gamma(\nu)}\Bigg(\sqrt{2\nu}\frac{|{\bf x}_i - {\bf x}_j| }{\rho}\Bigg)^\nu K_\nu\Bigg(\sqrt{2\nu}\frac{|{\bf x}_i - {\bf x}_j|}{\rho}\Bigg)
\end{equation}
where $\Gamma$ is the gamma function, $K_\nu$ is the modified Bessel function of the second kind, and $\rho$ and $\nu$ are non-negative globally constant hyperparameters which control spatial range and smoothness, respectively. A Mat\'ern kernel is $\nu - 1$ times differentiable and reduces to a Gaussian kernel in the limit $\nu \rightarrow \infty$ while becoming an exponential kernel when $\nu = 1/2$ \cite{Genton_2002}. It resultantly covers a wide class of kernels and confers flexibility. Nevertheless, the hyperparameters are quite restrictive if simply constants \cite{heinonen16,2d-gpr-tomography}. Therefore, a version of the generalized nonstationary Mat\'ern kernel is encoded for GP regression \cite{Paciorek_2006,Plagemann_2008}:
\begin{align}
\label{adaptive_kernel}
    k({\bf x}_i,{\bf x}_j) &= 
    \sigma^2_k \frac{2^{1-\nu}}{\Gamma(\nu)}
    \frac{|\rho_i|^{1/2}|\rho_j|^{1/2}}{\sqrt{\frac{1}{2}\rho^2_i + \frac{1}{2}\rho^2_j}} 
    \Bigg(2\sqrt{\frac{2\nu |{\bf x}_i - {\bf x}_j|^2}{\rho^2_i + \rho^2_j}}\Bigg)^\nu \nonumber
    \\ & \indent K_\nu\Bigg(2\sqrt{\frac{2\nu |{\bf x}_i - {\bf x}_j|^2}{\rho^2_i + \rho^2_j}}\Bigg)
\end{align}
where $\rho$ varies across the entire multidimensional domain and adapts to optimize the length scale based upon the experimental data being trained upon in each individual plasma discharge. The kernel accomplishes learning point estimates of local smoothness by representing the primary GP's locally isotropic length scale hyperparameter by a secondary GP with radial basis function (RBF) kernel that allows global variation of $\rho$ across the spatiotemporal grid. It is this second-level GP which introduces the notion of a deep process and adaptivity to the overall regression technique. A stationary kernel is purely a function of $\bf{x}_i - \bf{x}_j$, while additional local dependence exists in (\ref{adaptive_kernel}) through $\rho$ which introduces nonstationary behaviour \cite{Plagemann_2008}. A subset of experimental data is automatically selected with clustering centers determined via a k-means algorithm for training the secondary GP at these specific points. 

\begin{figure}
\centering
\includegraphics[width=0.75\linewidth]{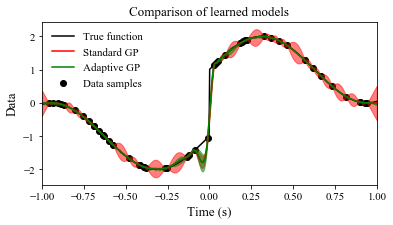} \\
\includegraphics[width=0.75\linewidth]{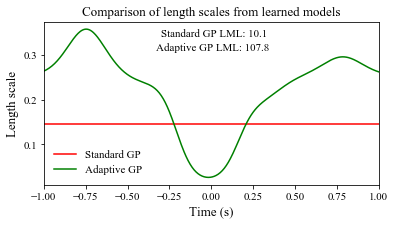}
\caption{Comparison of two separate GPs on 1-dimensional data: one employs a standard Mat\'ern kernel while the other includes an adaptive length scale. Fits to the original data samples (top) and computed length scales (bottom) are displayed courtesy of J.H. Metzen.}
\label{fig:1dlls}
\end{figure}
\begin{figure} 
\centering
\includegraphics[width=0.85\linewidth]{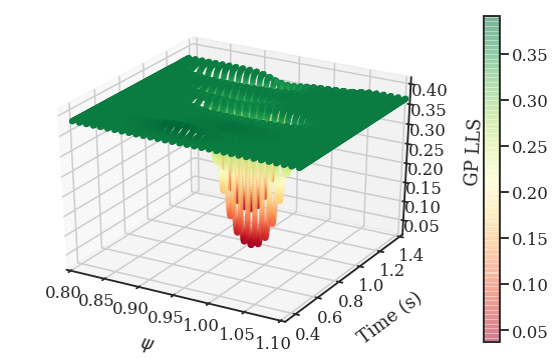} \\
\includegraphics[width=0.85\linewidth]{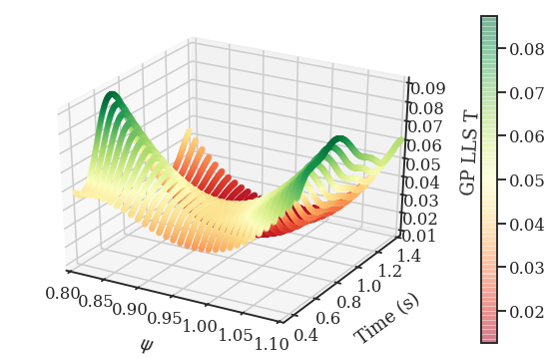}

\caption{Two examples of adaptively learning length scales that vary across the spatiotemporal domain by training identical GP models on experimentally measured electron density (top) and temperature (bottom) from the Thomson scattering plasma diagnostic.}
\label{fig:2dlls}
\end{figure}

Figure \ref{fig:1dlls} demonstrates a basic 1-dimensional example of a sinusoidal function with imposed discontinuity. The advantage conferred by the adaptive length scale can be quantitatively observed by comparing the log marginal likelihood (LML) \cite{Rasmussen_2005} between a standard Mat\'ern kernel and one with locally adaptive length scale. The order of magnitude improvement of LML (which is a logarithmic quantity) in Figure \ref{fig:1dlls} occurs because a stationary Mat\'ern kernel is forced to decrease its constant global length scale considerably while an adaptive length scale permits reducing its value locally only near the discontinuity. The adaptive length scales not only provide the capability to better capture singular or transient phenomena on otherwise slowly-varying profiles but importantly improves uncertainty estimates across the domain. We set $\nu = 3/2$ which allows for potentially stiff behaviour and this value can be modified in the source code, if sought, and kept as a variable for optimization to capture an entire spectrum of kernel functions. The user can freely specify upper and lower bounds on length scales to be learned across the grid which are given uniform prior distributions during training. 
As a preview for the application of these methods, the length scales can be extended to multidimensional scenarios as depicted in Figure \ref{fig:2dlls} where the learned input length scale varies across the entire spatial and temporal domain. The exact same kernel was initialized for both the top and bottom data sets used in Figure \ref{fig:2dlls}, but since the datasets were themselves different, the locally learned length scales vary. A custom stochastic optimizer based on differential evolution is used in these examples for GP hyperparameter-tuning and finally polished off with gradient-based descent. Finding a global optimum in the likelihood function is not guaranteed, therefore this is helpful because the loss function is highly multimodal and generally challenging for simple gradient-based methods acting on non-convex problems.  


    
\subsection{Heteroscedasticity}
\begin{figure}
\centering
\includegraphics[width=0.85\linewidth]{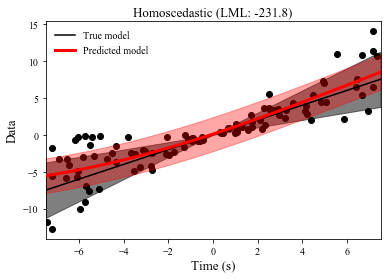} 
\\
\includegraphics[width=0.85\linewidth]{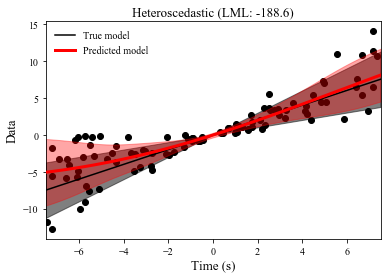}
\caption{Comparison of two separate GPs on 1-dimensional data both using an RBF kernel but one utilizes an additional homoscedastic component (top) and the other a heteroscedastic component (bottom) in the full covariance structure. Courtesy of J.H. Metzen.}
\label{hetero_figure}
\end{figure}

Heteroscedasticity in this context refers to learning intrinsic scatter associated with the noisy variable, $y$, and introducing non-constant variances. The full covariance function applied in the GP can be broken down into an adaptive kernel and noisy variance component:

\begin{equation}
    \Sigma({\bf x}_i,{\bf x}_j) = \underbrace{k({\bf x}_i,{\bf x}_j)}_{\text{adaptive}} + \underbrace{\sigma^2_n({\bf x}_i) \delta_{ij}}_{\text{heteroscedastic}}
\end{equation}
and heteroscedasticity is mathematically defined by $\sigma_n({\bf x}_i)$ having an explicit dependence on points in the input space. To contrast, homoscedasticity would entail a globally constant $\sigma_n$. To more vividly demonstrate the benefit of heteroscedasticity, a 1-dimensional example is displayed in Figure \ref{hetero_figure} by applying both homoscedastic and heteroscedastic components. The function to be learned is a linear relationship with variance growing quadratically along the abscissa. In this scenario with a homoscedastic noise model, the GP is forced to learn a constant intrinsic scatter in the underlying data commonly represented by white noise functions of constant amplitude. It is evident that enabling a heteroscedastic covariance function better captures the distribution of observed data. The mean estimates of both models are equivalent, but the predicted variances across the domain are markedly different and the heteroscedastic example obtains a larger LML and better captures intrinsic scatter in the data. The non-constant variances are learned across the domain using a k-means algorithm to once again identify clustering centers to provide characteristic noise estimates even when explicit error bars are absent. These prototype values are then extended across the domain by using a pairwise RBF kernel. Both the adaptive length scale kernel and heteroscedastic components are combined to significantly improve overall fitting and stability of the numerical optimization, e.g. avoid zero variances while training. This heteroscedastic term in the full data-driven covariance function is modular to an extent and can be optionally subtracted away to output confidence intervals ($\mathbb{V}[f_*]$) instead of prediction intervals ($\mathbb{V}[y_*]$) which are wider and account for intrinsic scatter in observations. (Note that $\mathbb{E}[f_*] \equiv \mathbb{E}[y_*]$.) 

\subsection{Multidimensional}

GPs do not require a fixed discretization scheme over the physical domain and hence easily handle spatially and temporally scattered diagnostic measurements involved in nonlinear regression. The generalized kernel above is encoded to handle data spanning any number of dimensions. Due to the highly non-convex optimization problem associated with finding optimal hyperparameters across the multidimensional space, the training applies stochastic optimization (Storn algorithm) with gradient descent at the end to help avoid becoming trapped in local minima. An associated error checking routine has been developed and available in the code on GitHub to automatically identify regions of the domain, if any, over which the GP did not successfully converge during training and requiring further optimization. Generally, multidimensional sampling of the GP is performed by applying 
\begin{equation}
    f_* = \mu_* + B\mathcal{N}(0,I)
\end{equation}
where $B B^T = \Sigma_*$ and $B$ is a triangular matrix known as the Cholesky decomposition of the covariance matrix. To account for particular constraints such as monotonicity or positivity with respect to the mean in sampled profiles, modified or truncated normal distributions can be enabled in the code to permit constrained GP sampling to yield physically relevant results. Finally, the technique's flexibility allows it to automatically handle vastly different data sets (e.g. thousands of separate discharges fitted in parallel) to construct large scale multidimensional profile databases with minimal user input.

\section{APPLICATION TO EXPERIMENTAL DATA}
\begin{figure}
\centering
\includegraphics[width=0.95\linewidth]{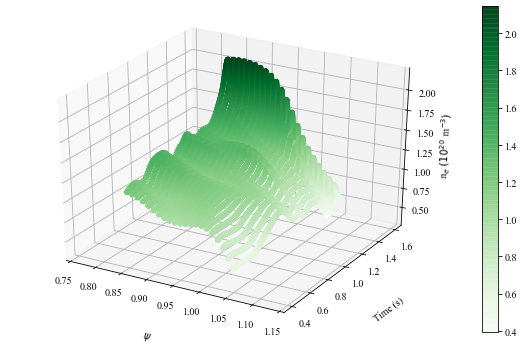} 
\includegraphics[width=0.95\linewidth]{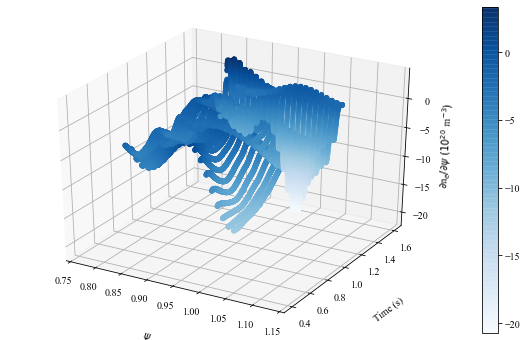}
\caption{Electron density and spatial gradients fitted by the adaptive heteroscedastic GP accounting for both spatial and temporal evolution of experimental data across the entire discharge over the edge-pedestal region (i.e. $0.85 < \psi < 1.05$, where $\psi$ is the normalized poloidal flux coordinate and applied for tracking data on both open and closed field lines).}  
\label{2D-GPR}
\end{figure}

The aforementioned adaptive heteroscedastic GP is now directly applied on experimental data originating from the Thomson scattering diagnostic on Alcator C-Mod which consists of two Nd:Yag lasers, each pulsing at 30 Hz with an approximate spatial resolution of 1.0 cm and 1.3 mm in the core and edge, respectively \cite{JWHughes-TS-diagnostic}. Accordingly, we can spatially resolve electron dynamics on scales of the poloidal ion gyroradius. Shot 1091016033 is analyzed which is partly displayed with time-averaged profiles in Figure \ref{Whyte2010_LHI} and exhibits L-, H-, and I-mode behaviours within this single discharge as detailed in \cite{Whyte_2010}. Ion cyclotron range of frequencies (ICRF) heating of up to 5 MW is applied in the experiment with power primarily deposited in the core on the hydrogen minority species. The on-axis magnetic field is 5.6 T with a plasma current of approximately 1.2 MA. There is a wide range of plasma behaviour associated with time-varying density and temperature pedestal structure even in this single discharge including transport barrier formation and confinement mode transitions necessitating a suitably robust regression method. The tools outlined above are therefore now demonstrated on shot 1091016033, and can be easily automated for edge data analyses across any set of discharges.
\begin{figure}
\begin{center}
    \includegraphics[width=0.975\linewidth]{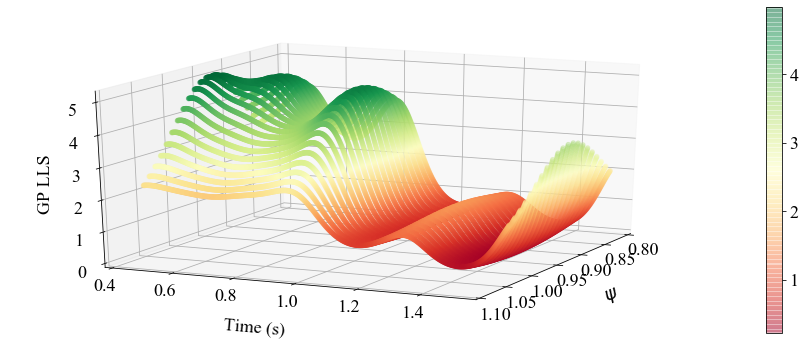}  
\caption{Corresponding length scales, $\rho$, learned across the spatial and temporal domain based on the adaptive heteroscedastic GP training upon spatiotemporally evolving data.} 
\label{lls_n_2d}
\end{center}
\end{figure} 
\begin{center}
\begin{figure*}  
\includegraphics[width=0.325\linewidth]{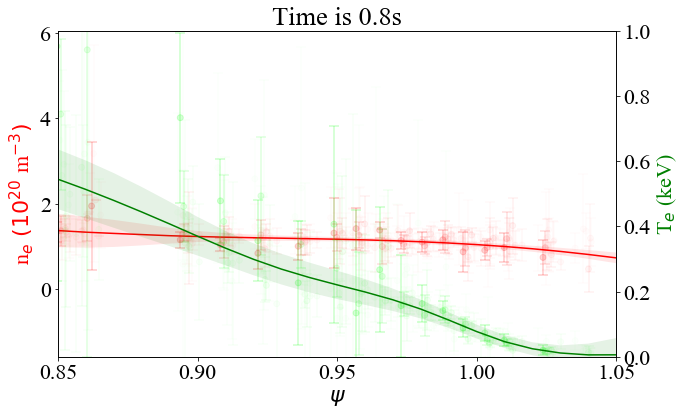} 
\includegraphics[width=0.325\linewidth]{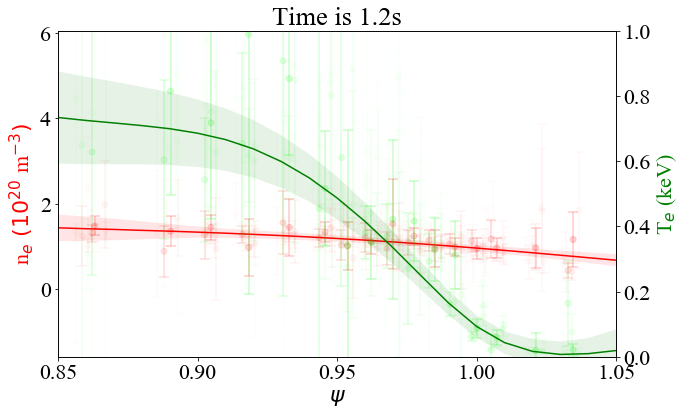}
\includegraphics[width=0.325\linewidth]{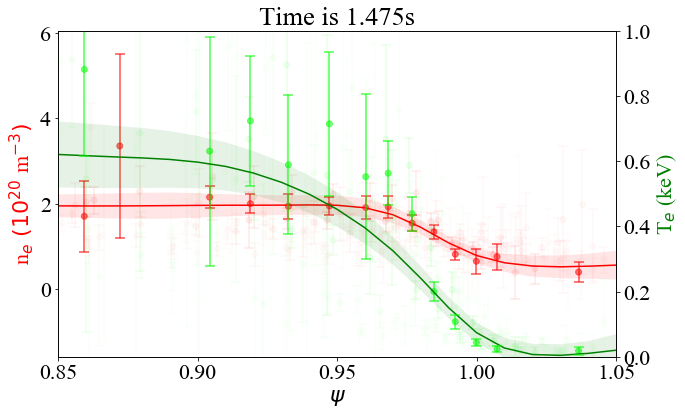}
\\ 
\includegraphics[width=0.325\linewidth]{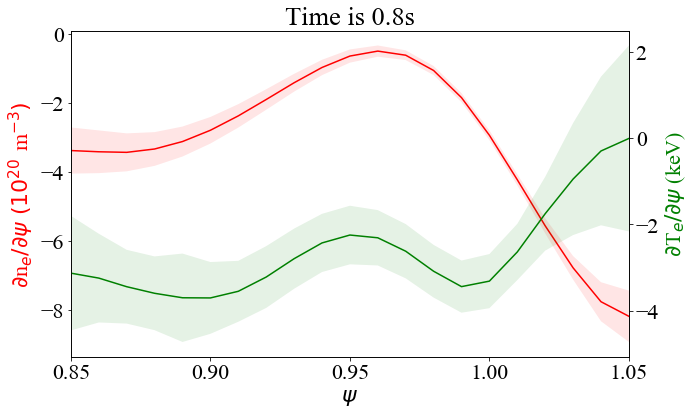} 
\includegraphics[width=0.325\linewidth]{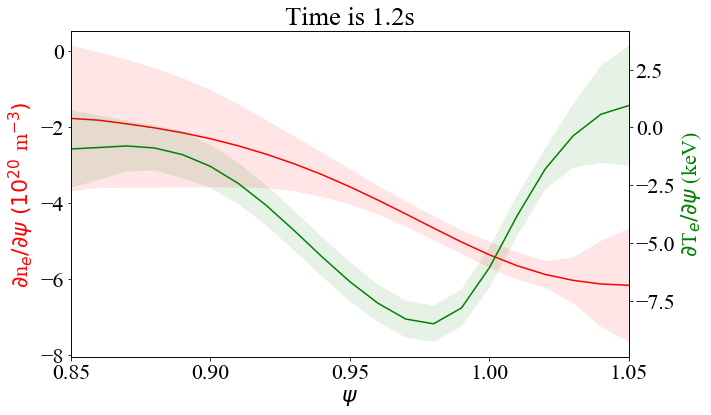} 
\includegraphics[width=0.325\linewidth]{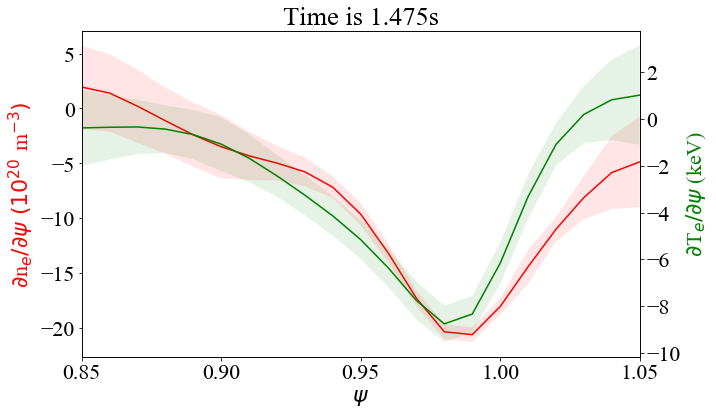}
\caption{Electron density and temperature measurements during  L- (left), I- (middle), and H-modes (right) fitted by the adaptive heteroscedastic GP without time-averaging experimental data. Proximity of experimental data to the time at which the GP is predicting is indicated by transparency of the data points. Note that 95\% prediction intervals are displayed for the top three plots while 95\% confidence intervals are applied for the bottom three plots.}
\label{1D-GPR}
\end{figure*}  
\end{center}
To begin, the trained GPs can compute expected mean density and temperature along with corresponding uncertainties across the entire spatiotemporal domain covered by the plasma diagnostic. Spatial gradients (or time derivatives) can be directly produced and are displayed in Figure \ref{2D-GPR} for electron density across the grid as well. The key advantage of applying the adaptive GP in this scenario is its ability to learn spatiotemporal correlations across the entire domain without filtering data. This freedom in training is evident in the learned variable length scales for density across the discharge as visualized in Figure \ref{lls_n_2d}. Particular time slices can also be evaluated by the GPs. Fitted L-, I-, and H-mode profiles from shot 1091016033 are displayed in Figure \ref{1D-GPR} without employing any time-averaging or filtering of experimental data as required when repeatedly applying a modified tanh function as in Figure \ref{Whyte2010_LHI}, which can miss important profile variation even within a confinement regime (e.g. while ramping up ICRF heating). Furthermore, density and temperature gradients along with uncertainties for all quantities are well-defined across the experiment during nominal L-, H-, and I-mode phases corresponding to times of 800, 1200, and 1475 milliseconds, respectively \cite{Whyte_2010}. 
A major benefit derived from applying the adaptive heteroscedastic GP is its ability to provide defining features (e.g. spatial and temporal gradients) of experimental profiles automatically across entire discharges. In past classification exercises to develop large confinement regime databases \cite{Mathews_2019}, individual time slices needed to be manually reviewed for the presence of density and/or temperature pedestals to identify windows containing different regimes (e.g. L-, H-, or I-modes). This is a highly time-intensive and arguably subjective route to develop meaningful confinement regime databases since discretely classifying plasma behaviour can miss underlying nuances (e.g. staircase pedestals, profile hollowing). Therefore, applying this multidimensional GP regression method permits automating characterization of profiles based upon certain observed quantitative features of confinement regimes such as gradient scale lengths. Namely, it can handle pertinent dynamics across the edge-pedestal region where properties atop the pedestal are expected to be different from relevant spatiotemporal scales in the relatively colder SOL when crossing the separatrix. Different varieties of stationary confinement regimes may have quite different profile dynamics and structure for temperature and density necessitating adaptive fitting capability. Additionally, accounting for temporal evolution of plasma profiles helps capture subtleties in types of transitions which are essential to better understand how confinement regimes evolve since no two modes nor transitions are necessarily identical, especially with hysteresis being prevalent across these bifurcation events. Resolving these features helps improve reconstructions of experimental measurements for further large scale analysis whether in stability codes or scientific machine learning or characterizing upstream conditions of background plasmas for divertor heat flux studies \cite{Silvagni_2020}. For example, the multidimensional GP can provide key profile information into plasma turbulence simulations (e.g. inputs for global gyrokinetic codes or comparisons with EPED \cite{Snyder_2009}) which may require sampling inputs such as gradient profiles to output sufficient statistics. Overall, this regression tool can be deployed for automatically providing multidimensional density and temperature profiles involving few sharp features for numerical analysis purposes. While the GP does not explicitly find L-, H-, or I-modes, it outputs time-dependent gradient profiles and corresponding uncertainties to systematically scan characteristics of confinement regimes across entire discharges such as identifying steepening in edge temperature gradients and formation of staircase pedestals. Structure imposed on learned edge gradient profiles is minimized by using a data-driven kernel function which can be critical to model plasmas near sensitive instability boundaries. Additionally, the denoised equilibrium plasma dynamics can be useful observational constraints in scientific learning applications \cite{Mathews2020_PRL}.

\section{Conclusion}
An adaptive multidimensional GP routine is outlined and utilized to automate fitting and uncertainty estimation of edge-pedestal measurements from the Thomson scattering diagnostic on the Alcator C-Mod tokamak while trying to be robust to edge-pedestal phenomena. Spatiotemporal evolution of plasma density and temperature is tracked with varying length scales extant in experimental data including the formation of both particle and energy transport barriers. The application is focused on edge measurements of tokamak plasmas with relevance to numerical analysis of the pedestal, but these techniques extend beyond analysis of the edge-pedestal region and can be suitably adapted to novel scenarios exhibiting singular transient events. The GP introduced provides an automated tool to tackle nonlinear multidimensional regression tasks and helps resolve measurements of equilibrium profiles with dynamics spanning a wide range of physical scales.

\section*{Acknowledgement}
We wish to give special credit and thanks to J.H. Metzen for original development of the code and package extensions created through \texttt{gp\_extras} (github.com/jmetzen/gp\_extras) and \texttt{scikit-learn}. This paper's examples can be found on Github (github.com/AbhilashMathews/gp\_extras\_applications) and the repository will become public upon publication. 

 \appendix
\setcounter{equation}{0}
\renewcommand{\theequation}{A.\arabic{equation}}

For completeness, a simple derivation of the GP machinery to perform conditional predictions from observed data, i.e. (\ref{GPR_machinery1}) and (\ref{GPR_machinery2}), based originally upon \cite{Tso2010} is reproduced below. As previously remarked, a normally distributed random variable, ${\bf y}$, in $N$-dimensions is modelled by 
\begin{equation}
P({ \bf y} \vert {\boldsymbol\mu}, { \bf \Sigma}) = \frac {1}{2\pi^{N/2} \lvert { \bf  \Sigma} \rvert^{1/2}} \exp[-\frac{1}{2}({ \bf y}-{\boldsymbol\mu})^T { \bf \Sigma}^{-1} ({ \bf y}-{\boldsymbol\mu})],
\end{equation}
where ${ \bf \Sigma}$ is a positive semi-definite covariance matrix, ${\boldsymbol\mu} = [\mu_1, ..., \mu_N]$, ${ \bf y} = [y_1, ..., y_N]$, and ${\bf \Sigma}_{i,j} = \mathbb{E}[(y_i - \mu_i)(y_j - \mu_j)]$. In this formalism, the conditional probability of predicting a new point (or set of points), ${\bf y}_b$, can be ascertained from an observed point (or set of points), ${ \bf y}_a$, through the posterior distribution: ${ \bf y}_b \vert { \bf y}_a \sim \mathcal{N}({\boldsymbol\mu}_{{ \bf y}_b \vert { \bf y}_a},{\Sigma}_{{ \bf y}_b \vert { \bf y}_a})$ \cite{Rasmussen_2005}. In a Bayesian framework, the multivariate normal distribution's conditional mean and variance are:
\begin{align}
{\boldsymbol\mu}_{{ \bf y}_b \vert { \bf y}_a} = \boldsymbol\mu_b+\Sigma_{b,a}{\Sigma_{a,a}}^{-1}({\boldsymbol y_a}-\boldsymbol\mu_a)
\end{align}
\begin{align}
{\Sigma}_{{ \bf y}_b \vert { \bf y}_a} = \Sigma_{b,b}-\Sigma_{b,a}{\Sigma_{a,a}}^{-1}\Sigma_{a,b}
\end{align}
 Following the treatment in \cite{Tso2010}, this can be derived by defining ${\bf z} \equiv {\bf y}_b + {\bf A} {\bf y}_a $ where ${\bf A} \equiv -\Sigma_{b,a} \Sigma^{-1}_{a,a}$, implying 
\begin{align} 
{\rm cov}({\bf z}, {\bf y}_a) &= {\rm cov}( {\bf y}_{b}, {\bf y}_a ) + 
{\rm cov}({\bf A}{\bf y}_a, {\bf y}_a) \nonumber \\
&= \Sigma_{b,a} + {\bf A} {\rm var}({\bf y}_a) \nonumber \\
&= \Sigma_{b,a} - \Sigma_{b,a} \Sigma^{-1}_{a,a} \Sigma_{a,a} \nonumber \\
&= 0
\end{align}
Consequently, ${\bf z}$ and ${\bf y}_a$ are uncorrelated and, since they are assumed jointly normal in GP regression, they are independent. It is evident $\mathbb{E}[{\bf z}] = {\boldsymbol \mu}_b + {\bf A}  {\boldsymbol \mu}_a$, and it follows that the conditional mean can be expressed as
\begin{align}
\mathbb{E}[{\bf y}_b | {\bf y}_a] &= \mathbb{E}[] {\bf z} - {\bf A} {\bf y}_a | {\bf y}_b] \nonumber \\
& = \mathbb{E}[{\bf z}|{\bf y}_a] -  \mathbb{E}[{\bf A}{\bf y}_a|{\bf y}_a] \nonumber \\
& = \mathbb{E}[{\bf z}] - {\bf A}{\bf y}_a \nonumber \\
& = {\boldsymbol \mu}_b + {\bf A}  ({\boldsymbol \mu}_a - {\bf y}_a) \nonumber \\
& = {\boldsymbol \mu}_b + \Sigma_{b,a} \Sigma^{-1}_{a,a} ({\bf y}_a- {\boldsymbol \mu}_a)
\end{align}
which proves the conditional mean, and 
\begin{align}
{\rm var}({\bf x}-{\bf D}{\bf y}) &\equiv {\rm var}({\bf x}) + {\bf D}{\rm var}({\bf y}){\bf D}^T \nonumber \\ & - {\rm cov}({\bf x},{\bf y}){\bf D}^T - {\bf D}{\rm cov}({\bf y},{\bf x})
\end{align}
implies
\begin{align}
{\rm var}({\bf y}_b|{\bf y}_a) &= {\rm var}({\bf z} - {\bf A} {\bf y}_a | {\bf y}_a) \nonumber \\
&= {\rm var}({\bf z}|{\bf y}_a) + {\rm var}({\bf A} {\bf y}_a | {\bf y}_a) \nonumber \\ &- {\bf A}{\rm cov}({\bf y}_a, {\bf z}) - {\rm cov}({\bf z}, {\bf y}_a) {\bf A}^T \nonumber \\
&= {\rm var}({\bf z}|{\bf y}_a) = {\rm var}({\bf z})
\end{align}
Plugging the above result into the conditional variance,
\begin{align}
{\rm var}({\bf y}_b|{\bf y}_a) &= {\rm var}( {\bf y}_b + {\bf A} {\bf y}_a ) \nonumber \\
&= {\rm var}( {\bf y}_b ) + {\bf A} {\rm var}( {\bf y}_a ) {\bf A}^T + \nonumber \\ &{\bf A} {\rm cov}({\bf y}_b,{\bf y}_a) + {\rm cov}({\bf y}_a,{\bf y}_b) {\bf A}^T \nonumber \\
&= \Sigma_{b,b} +\Sigma_{a,b} \Sigma^{-1}_{a,a} \Sigma_{a,a}\Sigma^{-1}_{a,a}\Sigma_{a,b} \nonumber \\ &
\indent - 2 \Sigma_{b,a} \Sigma_{a,a}^{-1} \Sigma_{a,b} \nonumber \\
&= \Sigma_{b,b} +\Sigma_{b,a} \Sigma^{-1}_{a,a}\Sigma_{a,b}
\nonumber \\ &- \indent 2 \Sigma_{b,a} \Sigma_{a,a}^{-1} \Sigma_{a,b} \nonumber \\ 
&= \Sigma_{b,b} -\Sigma_{b,a} \Sigma^{-1}_{a,a}\Sigma_{a,b}
\end{align}
Therefore, all that is required for the GP machinery are priors over ${\bf y}_a$ and ${\bf y}_b$ to obtain $\boldsymbol\mu_a$ and $\boldsymbol\mu_b$, respectively, and a covariance function which is defined in this paper by a heteroscedastic uncertainty component, $\epsilon$, along with an adaptive kernel function, $k({\bf x}_i,{\bf x}_j)$. The resulting covariance, $\Sigma_{i,j} = k({\bf x}_i,{\bf x}_j) + \epsilon({\bf x}_i)  \delta_{i,j}$, describes similarity between data points and accounts for spatiotemporal correlations in the data since ${\bf x}$ represents the independent variables (e.g. $\psi$ and $t$). The hyperparameters of the prescribed adaptive kernel function are optimized through maximum {\it a posteriori} estimation on each individual discharge's observed data. Practically, given an experimental set of noisy measurements at arbitrary positions and times, this formalism allows us to infer expected values of these measurements across our spatiotemporal domain. 
\bibliographystyle{IEEEtran}
\bibliography{sample.bib}
\begin{IEEEbiographynophoto}{Abhilash Mathews}
is currently a Ph.D. candidate in the Department of Nuclear Science and Engineering, Massachusetts Institute of Technology, Cambridge, MA, USA.
\end{IEEEbiographynophoto}
\begin{IEEEbiographynophoto}{Jerry W. Hughes}
is a Principal Research Scientist at the MIT Plasma Science and Fusion Center in Cambridge, MA, USA.
\end{IEEEbiographynophoto}

\end{document}